\documentclass[3p,times,sort&compress]{elsarticle}
\usepackage{amsmath}
\usepackage{bm}
\usepackage{amssymb}
\usepackage{graphicx}
\usepackage{hyperref}
\usepackage{xspace}
\usepackage{bm}
\usepackage{slashed}
\usepackage{xspace}
\usepackage{color}

\usepackage{multirow}
\usepackage{booktabs}
\AtBeginDocument{
\heavyrulewidth=.08em
\lightrulewidth=.05em
\cmidrulewidth=.03em
\belowrulesep=.65ex
\belowbottomsep=0pt
\aboverulesep=.4ex
\abovetopsep=0pt
\cmidrulesep=\doublerulesep
\cmidrulekern=.5em
\defaultaddspace=.5em
}

\newcommand{\be}{\begin{equation}}
\newcommand{\ee}{\end{equation}}
\newcommand{\ba}{\begin{eqnarray}}
\newcommand{\ea}{\end{eqnarray}}

\newcommand{\Xb}{\ensuremath{X_{bs}}}

\usepackage{xcolor}
\definecolor{light-gray}{gray}{0.8}

\newcommand{\bbb}[1]{\boldsymbol{#1}}

\newcommand{\eg}{\emph{e.g.}\xspace}

\begin{document}

\title{ Hybridized Tetraquarks}

%\preprint{JLAB-THY-16-2232}

%\newcommand{\sapienza}{Dipartimento di Fisica, `Sapienza' Universit\`a di Roma \\
%P.le Aldo Moro 5, I-00185 Roma, Italy}
%\newcommand{\columbia}{Department of Physics, 538W 120th Street,
%Columbia University, New York, NY, 10027, USA}
%\newcommand{\jlab}{Theory Center, Thomas Jefferson National Accelerator Facility, \\ 12000 Jefferson Avenue,
%Newport News, VA 23606, USA}
%\newcommand{\cern}{CERN, Theory Division, Geneva 23, Switzerland}
%\newcommand{\desy}{Deutsches Elektronen-Synchrotron DESY, D-22607 Hamburg, Germany}
%\newcommand{\infn}{INFN Sezione di Roma, P.le Aldo Moro 5, I-00185 Roma, Italy}
%
%\author{A.~Esposito}
%\affiliation{\columbia}
%\author{A.~Pilloni}
%\affiliation{\jlab}
%\affiliation{\infn}
%\author{A.D.~Polosa}
%\affiliation{\sapienza}
%\affiliation{\infn}
%\affiliation{\cern}

\author[uno]{A. Esposito}
\author[due,cinque]{A. Pilloni}
\author[tre,cinque,quattro]{A.D. Polosa}
\address[uno]{Department of Physics, 538W 120th Street, Columbia University, New York, NY 10027, USA}
\address[due]{Theory  Center,  Thomas  Jefferson  National  Accelerator  Facility, 
12000  Jefferson  Avenue,  Newport  News,  VA  23606, USA}
\address[cinque]{INFN Sezione di Roma,
P.le Aldo Moro 5, I-00185 Roma, Italy}
\address[tre]{Dipartimento di Fisica, ``Sapienza'' Universit\`a di Roma, P.le A. Moro 2, I-00185 Roma, Italy}
\address[quattro]{CERN, Theory Division, Geneva 1211, Switzerland}

\begin{abstract}
We propose a new interpretation of the neutral and charged $X,Z$ exotic hadron resonances. \mbox{\emph{Hybridized}-tetraquarks} are neither purely compact tetraquark states nor bound or loosely bound molecules but rather a manifestation of the interplay between the two. While meson molecules need a negative or zero binding energy, its counterpart for \mbox{$h\,$-tetraquarks} is required to be positive.  The formation mechanism of this new class of hadrons is inspired by that of Feshbach metastable states in atomic physics. The recent claim of an exotic resonance in the $B_s^0\, \pi^\pm$ channel by the D0 collaboration and the negative result presented subsequently by the LHCb collaboration are understood in this scheme, together with a considerable portion of  available data on $X,Z$ particles. Considerations on a state with the same quantum numbers as the $X(5568)$ are also made.
 \end{abstract}

\begin{keyword}
Tetraquarks, Exotic Hadrons, Feshbach Resonances
\PACS 12.39.Mk, 12.39.-x, 12.40.Yx\\
\href{http://arxiv.org/1603.07667}{arXiv:1603.07667}, JLAB-THY-16-2232
\end{keyword}

\maketitle

%{\bf\emph{ Introduction.}}
\section{Introduction}
The two most popular phenomenological models introduced to describe the $XYZ$ resonances are the compact tetraquark~\cite{tetra1,tetra3,lebed,rv} and the loosely bound meson molecule~\cite{mols}. While in the first description the exotic mesons are four-quark objects tightly bound by color forces, in the second one they are real bound states in a shallow inter-hadron potential (for a review, see~\cite{review}).

In this Letter we propose a new interpretation of these states: $h\,$-tetraquarks\footnote{Note that our states do not have any relation to the  gluonic hybrids~\cite{hybrids}. 
The term `hybridization' here is taken from the physics of Feshbach resonances in cold atom systems, with the meaning explained in the text.} result from an hybridization 
between the discrete levels of the tetraquark potential and the levels of the continuous spectrum of the 
two-meson potential.

The guiding principle to identify $h\,$-tetraquarks is to first write the diquark composition of the would-be-compact tetraquarks along the lines described in~\cite{tetra1,tetra3}. This gives an estimate of the energy of the discrete level of interest. The strongly bound diquark-antidiquark state  can be Fierz rearranged in a number of color singlet pairs which can be of the form hidden-flavor~+~light meson or two open flavor mesons, having quantum numbers compatible with the initial tetraquark state. The spin of the light quark component is allowed to flip, whereas the spin of the heavy quark pair has to be the same in both the compact tetraquark and  the meson pair description.

The mass of the would-be-compact tetraquark is computed with the methods of~\cite{tetra1,tetra3}. The masses of the corresponding different would-be-hadron molecules are computed as sums of the masses of the components of the pair, with no interaction energy. A reference molecule is taken in our analysis: the one having the closest possible mass, {\it from below}, to the tetraquark (diquark-antidiquark) discrete level.

The meson pair is allowed to interact in the continuous spectrum of some unknown shallow meson-meson potential which is assumed not to have (negative-energy) bound states.
A level in the continuous spectrum of the two-body system and the near discrete level of the compact tetraquark can match as illustrated below.
If this matching is realized, a sort of {\it `hybridization'} of the hadron molecule into the compact structure occurs.
 
The hybridized state is unstable as it can dissociate back into its free components ---  this is expected to be the major contribution to the width of the ground state tetraquarks. Other, less frequent, dissociation channels are also possible and partly contribute to the total width.  

The scattering in the continuous spectrum is assumed to be essentially inelastic because of the hybridization of the tetraquark final state, which produces a temporary rearrangement of the internal structure of the system.

What is important is that the {\it detuning} $\delta$, {\it i.e.} the distance in energy between the expected tetraquark discrete level (which we can estimate) and the onset of the continuous spectrum starting from the closest molecular threshold is {\it positive} and {\it small} with respect to the coupling $|\varkappa|$ responsible of the hybridization process.

 The case in which $\delta<0$ suggests a repulsion in the meson-meson channel, which is incompatible with hybridization. This might be the reason which forbids the charged partners of the $X(3872)$ and provides isospin violation (the charged threshold happens to be $4$~MeV above the tetraquark level).

\section{`Hybridization'}
\label{sec:hybrid}
For any given threshold, the scattering length $a$ in the open channel $P$ of the meson pair gets enhanced if a discrete level of the closed channel $Q$ of compact diquarkonia happens to be \emph{above} and close to the onset of the continuous spectrum of the pair, according to
\begin{subequations} \label{ascattering}
\begin{gather}
 a \simeq a_P - C\sum_n \frac{\langle \Psi_{ \alpha} |H_{PQ}|\Psi_n  \rangle\langle \Psi_n|H_{QP}|\Psi_{ \alpha}\rangle }{E_n-E_\alpha}\simeq \left(1-\frac{\varkappa}{\delta-E+i\epsilon}\right)a_P  \label{eq:overlap} \\
\text{Im}\,a  \sim \delta(E-\delta)\,\varkappa\, a_P
\label{eq:scatt},
\end{gather}  
\end{subequations}
where $|\Psi_{n}\rangle$ is the discrete level in the closed channel and $|\Psi_{ \alpha}\rangle$ is 
a continuous spectrum state above one of the thresholds  $\psi^{(\prime)}\pi,\eta_c\,\rho,\bar D^*D^*, \bar D^*D$, taking the $Z_c$ resonance as an example. The energy associated with $|\Psi_\alpha\rangle$ is $E_\alpha=E$ for brevity.  $a_P$ represents the small scattering length at zero coupling between open and closed channels. $C$ is a positive numerical constant, $H_{QP}$ is the non-hermitian Hamiltonian which couples the closed and open channels. In \eqref{eq:overlap}, $\delta=E_n-E_\text{thr}$ is the small `detuning' between the discrete level and the closest threshold (onset of the continuous spectrum). 
The effective coupling $\varkappa$ is real and depends on the overlap integrals in \eqref{eq:overlap}. 
It contributes to the inelastic channel in which a compact tetraquark is formed as a metastable state (the inverse process also occurs).  

The phenomenon described induces a resonant enhancement in the production of $h\,$-tetraquarks and is compatible with their production in high energy and high transverse momentum proton-(anti)proton collisions, as opposed to what expected for real loosely bound molecules, as discussed in the literature~\cite{prod2,noifesh2}.

The inelastic cross section at low energy in the continuous spectrum of the open channel is (neglecting numerical constants)
\begin{equation}
\sigma_\text{in}\sim \frac{|\mathrm{Im}\, a|}{p},
\end{equation}
where $p$ is the relative momentum in the center-of-mass of the pair. Therefore, the rate at which the $h$-tetraquark is formed is
\begin{equation}
d\Gamma\sim \rho\, v\, \sigma_\text{in}\sim \delta(E-\delta)\,\left| \varkappa\, a_P\right|\frac{\rho}{m}
\label{uno}
\end{equation}
where $\rho$ is the density of initial states
\begin{equation}
\rho\sim d^3 p = (2m)^{3/2} \sqrt{E} \, dE.
\label{due}
\end{equation}
and the integral in $E$ is extended over some  $[0,E_{\rm max}]$ range. The Dirac-delta in~(\ref{uno}) gives an integral different from zero only if the detuning $\delta$ falls within the integration range, $\delta<E_{\rm max}$. In that case the level matching condition $E\sim\delta$ enhances the hybridization  of the would-be-molecule with the corresponding diquarkonium. Inserting~\eqref{due} into~\eqref{uno} gives~\footnote{Compare with the discussion in~\cite{bec}.}
\be
\Gamma\sim (2m)^{1/2}\, \left| \varkappa\, a_P\right| \, \sqrt{\delta}\sim A\,\sqrt{\delta}
\label{main}
\ee
It is difficult to estimate $E_{\rm max}$, the maximum relative energy in a would-be-molecule. In our view the hadronization state is a superposition of a diquarkonium state plus all possible molecular states allowed by quark flavors and quantum numbers.  We might reasonably expect that being the  color force screened between the color singlets components of the molecule, the relative energy must be smaller that what one would find in a compact system.

The total width of the state can be expressed as a sum on all available open-channels
\begin{equation}
\Gamma\sim \sum_{i} \Gamma_i.
\end{equation}
It is essential here to note that if only pure phase space were to be considered~\cite{wang}, then the open channel with the largest detuning would be the dominant one. On the contrary, the considerations made above show how the enhancement in Eq.~\eqref{ascattering} leads to select the \emph{closest} threshold (from below): all partial widths will be negligible, exception made for the one relative to a $\delta$ within the $[0,E_{\rm max}]$ integration range. This mechanism is inspired by the formation of metastable Feshbach states in atomic physics~\cite{bec} (in $XYZ$ context see~\cite{noifesh,noifesh2}, and in general in the strong interaction~\cite{fesh2}).

\begin{figure}[t]
\centering
\includegraphics[width=.5\columnwidth]{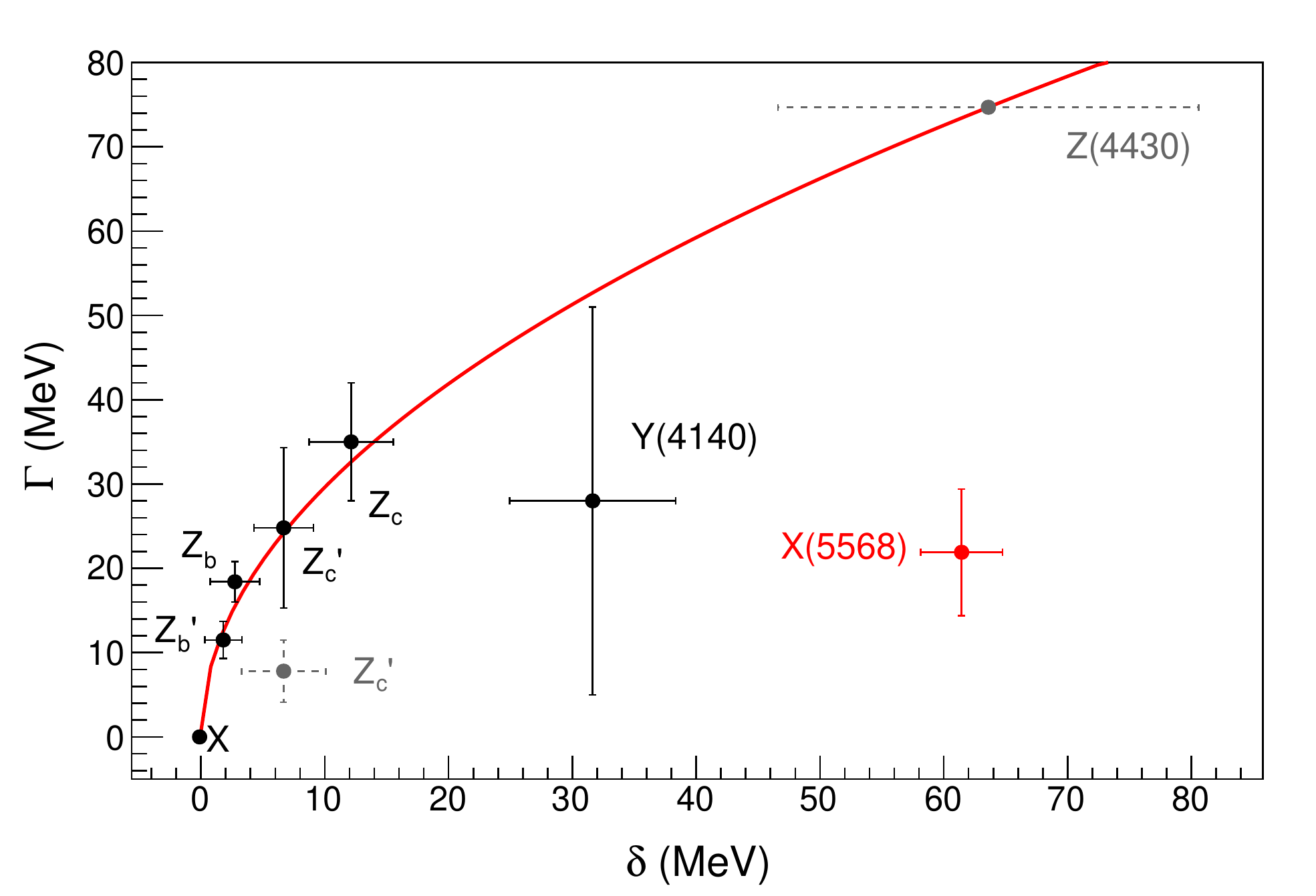}
\caption{Width of the observed exotic mesons as a function of their detuning, $\delta$ to the closest, from below, two-meson thresholds. The red point corresponds to the $X(5568)$ state whose observation has been claimed by D0. The solid curve is the fitted function $A \sqrt{\delta}$, with $A = 9.4 \pm 1.5~\text{MeV}^{1/2}$ with $\chi^2/\text{DOF} = 1.4 / 5$ (without the $Y(4140)$~\cite{cms}, the quality of the fit would be $\chi^2/\text{DOF} = 0.2 / 4$). The two points associated with the $Z_c^\prime(4020)$ correspond to the two measurements of its width obtained from $\bar D^{*0}D^{*+}$ (solid black) and the $h_c\pi$ (dashed gray) channels and which differ at $2\sigma$ level from each other. In the fit we considered the $\Gamma(Z_c^\prime) \simeq 25$~MeV measured in the $\bar D^{*0}D^{*+}$ channel.  We also show the prediction for the $Z(4430)$ width, which underestimates the total width as expected.} \label{fit_fesh}
\end{figure}

Therefore we now only consider the closest molecular channel, and for practical purposes use $\Gamma=A\sqrt{\delta}$ in place of Eq.~(\ref{main}). We observe that the widths and detunings in a broad class of observed resonances strictly obey this law with a common value for the $A$ parameter --- this can be appreciated by the very good fit in  Fig.~\ref{fit_fesh}.
 The fact that all data can be fitted with the same proportionality constant $A$, strongly supports for the described states to share the same nature. It also shows that this is not just a phase space effect. It is worth noting that Fig.~\ref{fit_fesh} implies that the $\varkappa$ coupling in~(\ref{main}) has to nearly cancel (within the errors of the fit) the $\sqrt{m}$ dependency on the reduced mass of the molecule. 

We might expect small variations among  different $a_P$'s. For example, open charm meson pairs have scattering lengths, $a_P$, likely larger than the ones for charmonium~+~light meson pairs. This might explain why the $Y(4140)$, which matches a $J/\psi\,\phi$ threshold, is slightly off in the description of Fig.~\ref{fit_fesh}, even though by merely a 1$\sigma$ deviation.     

The above arguments do not straightforwardly generalize to excited tetraquarks. In that case the closed channel is itself not stable against a de-excitation into its allowed tetraquark ground state. It then follows that the width predicted with the approach explained  will for sure underestimate the actual width of the state. For example, we consider the $Z(4430)$ which, in the tetraquark model, is the radial excitation of the $Z_c(3900)$~\cite{tetra3}. The closest threshold from below would be $\eta_c(2S)\,\rho$ with a detuning $\delta\simeq65$ MeV. The latter one is probably rather large to consider Eq.~\eqref{ascattering} without including  other discrete levels. Nevertheless the  width obtainedwould be $\Gamma\simeq80$ MeV which naturally underestimates the experimental one, $\Gamma=181\pm31$ MeV. Similarly, for the moment we do not extend the analysis to pentaquarks, being the experimental information not sufficient. In this case, not only one of the two observed resonances would be an excited state, but we do not have any hint about $A$. We indeed expect substantial differences from the present case since the baryon-meson scattering is different.

From Fig.~\ref{fit_fesh}  we  see that the ground states nicely fit into the $h\,$-tetraquarks pattern. In Table~\ref{table} we report the thresholds considered for each state.

\begin{table}[t!]
\begin{center}
 \begin{tabular}{l|c|c|c|c}
 & Thr. & $\delta$ & $A\sqrt{\delta}$ & $\Gamma$  \\ \hline
$X(3872)$ & $\bar D^0 D^{*0}$ & $0^\dagger$ & $0^\dagger$ & $0^\dagger$   \\
$Z_c(3900)$ & $\bar D^0 D^{*+}$ & $12.1$ & $34.8$ & $35.0$   \\ 
$Z_c^\prime(4020)$ & $\bar D^{*0} D^{*+}$ & $6.7$ & $25.9$ & $24.8^\P$   \\ 
$Y(4140)$ & $J/\psi\,\phi$ & $31.6$ & $52.7$ & $28.0$ \\
$Z_b(10610)$ & $\bar B^0 B^{*+}$ & $2.7$ & $16.6$ & $18.4$ \\ 
$Z_b^\prime(10650)$ & $\bar B^{*0} B^{*+}$ & $1.8$ & $13.4$ & $11.5$ \\  \hline
$X(5568)$ & $B_s^0\,\pi^+$ & $61.4$ & $78.4$ & $21.9$ \\ \hline
$\Xb$ & $B^+\bar K^0$ & 5.8 & 24.1 & ---
 \end{tabular}
 \end{center}
\caption{Comparison of the results obtained assuming a Feshbach mechanism at work with observed data. We find good agreement with the predicted widths. On the other, the $X(5568)$ observed by D0 has a width way smaller than the expected one. We also show the prediction for the $\Xb$ as explained in the text. For the $Y(4140)$, we considered the most recent result by CMS~\cite{cms}. $^\dagger$The mass of the $X(3872)$ is compatible with the value of the $\bar D^0D^{*0}$ threshold within errors, so we assume for the state to be slightly above it. $^\P$We show the value of the width of the $Z_c^\prime(4020)$ measured in the $\bar D^{*0} D^{*+}$ channel, which is $2\sigma$ away from the one measured in the $h_c \pi$ channel. }\label{table}
\end{table}

\section{The $X(5568)$}
The D0 experiment recently claimed the observation of a new narrow structure in the $B_s^0\,\pi^\pm$ invariant mass\footnote{Hereafter the charged conjugated modes --- \emph{e.g.} $\bar B_s^0\pi^\mp$ --- are understood.}~\cite{D0:2016mwd} --- dubbed $X(5568)$ ---, which promptly attracted some consideration~\cite{others}. The resonance parameters are given by $M=5567.8$~MeV and $\Gamma=21.9$~MeV. 
However, a preliminary analysis subsequently performed by the LHCb collaboration, on a $B_s^0$ sample 20 times larger than the D0 one, found no evidence of $X(5568)$~\cite{moriond}.  

As shown by the red dot in Fig.~\ref{fit_fesh}, the $X(5568)$ significantly deviates from the expectations of the previous section. The discrepancy is even more relevant if one considers that \emph{(a)} the interaction of a $B_s$ with a Goldstone boson should likely have a larger $a_P$ and \emph{(b)} the phase space of a light meson would lead to a steeper curve ($\Gamma\propto\delta$ in the chiral limit). 

\section{The $\Xb$ in the diquarkonium picture.}
 The D0 state would be unambiguously composed by four valence quarks with different flavors --- \mbox{$u$, $d$, $s$, $b$}. In principle, structures of this type might be accommodated in the tetraquark model~\cite{review}, see \eg \cite{tantaloni}. 

Within the constituent quark model the color-spin Hamiltonian describing the interaction between the different constituents of a hadron takes the following form
\begin{align} \label{H}
H=\sum_i m_i+2\sum_{i<j}\kappa_{ij}\bbb{S}_i\cdot \bbb{S}_j,
\end{align}
where $m_i$ are the masses of the constituents, $\bbb{S}_i$ their spins and $\kappa_{ij}$ some effective, representation-dependent chromomagnetic couplings. The spin-spin interaction is here understood to be a contact one.

In the diquark-antidiquark picture, the $X(5568)$ would be given by $[\bar b\bar q]_{{\bbb{3}}_c}[ s q^\prime]_{\bar{\bbb{3}}_c}$, with $q\neq q^\prime =u,d $. In the most recent and most successful type-II tetraquark model~\cite{tetra3}, given the spatial separation between the diquarks, the only relevant interaction is assumed to be the one between the spins within the tightly bound diquarks, meaning that every effective coupling is set to zero, except for $\kappa_{bq}$ and $\kappa_{sq^\prime}$. 

Given the small available phase space it is likely for the $X(5568)$ to be a scalar and therefore we will consider the ground $S$-wave state $\Xb\equiv[\bar b\bar q]_{S=0}[ s q^\prime]_{S=0}$, $S$ being the spin of the diquark. Its mass would then be given by
\begin{align}
M(\Xb)&=m_{[bq]}+m_{[sq]}+2\kappa_{bq}\,\bbb{S}_{\bar b}\cdot\bbb{S}_{\bar q}+2\kappa_{sq}\,\bbb{S}_{s}\cdot\bbb{S}_{q^\prime}\notag\\
&=  m_{[bq]}+\left(m_{[sq]}-3/2\, \kappa_{sq}\right) -3/2\, \kappa_{bq}.
\end{align}

The combination in parentheses corresponds to $m_{a_0}/2=\left(m_{[sd]}-3/2\, \kappa_{sq}\right)$, half the  mass of the scalar meson $a_0(980)$, as computed in~\cite{tetra1}. 
The parameters $m_{[bq]}$ and $\kappa_{bq}$, can be estimated starting from the masses of the observed $Z_b(10610)$ and $Z_b^\prime(10650)$ obtaining
\begin{align}
m_{[bq]}\simeq5315\text{ MeV};&\quad\kappa_{bq}\simeq22.5\text{ MeV}.
\end{align}
Indeed, using the type-II model~\cite{tetra3}, it is readily found that 
\begin{subequations}
\begin{align}
m_{[bq]}&=\frac{M(Z_b^\prime)+M(Z_b)}{4}, \\
\kappa_{bq}&=\frac{M(Z_b^\prime)-M(Z_b)}{2}.
\end{align}
\end{subequations}

With all the quantities at hand, the expected mass for the $\Xb$ state would be
\begin{align}
M(\Xb)\simeq5771\text{ MeV}.
\end{align}
As one can see this would put it very close to the $BK$ threshold, at $5778$~MeV. In this below-threshold situation, the hybridization  mechanism is expected to enhance the repulsion in the open channel, and the state will {\it not} be formed. However, the actual mass of the state could slightly deviate from the diquarkonium rough estimate: 
the diquarkonium level might happen to be right \emph{above} the  $BK$ threshold. In this case we expect to see a resonance  with a dominant $BK$ decay mode, to be observed in prompt $pp(\bar p)$ production, or in the $B_c$ decays. 

So, we can give a rough estimate of the expected width for the $\Xb$. Assuming the detuning to be of the same order as the one observed for the other ground state tetraquarks ($\delta\simeq6$ MeV) one predicts the $\Xb$ width to be roughly $\Gamma_{\Xb}\simeq20-30$ MeV. 

\section{Conclusions} 
We presented a new scheme to interpret in a consistent way quite a large portion of the best experimentally  assessed charged and neutral exotic hadron resonances. In this scheme the state very recently claimed by the D0 collaboration in the $B_s\pi^{\pm}$ channel does {\it not} fit in a remarkable way. The recent negative result reported by the LHCb collaboration on the non-existence of this resonance agrees with the expectations discussed here. Moreover we find that, if a state with the quantum numbers of the resonance observed by D0 actually exists, it has to be found slightly above the $BK$ threshold. The closed channel level predicted by the compact tetraquark Hamiltonian is actually computed at a position  a little below the $BK$ threshold. If the theoretical error in that estimate is less than we might expect, {\it no} resonance will be observed close to $BK$: hadron molecules (with zero or negative binding energy) are not formed  in large energy hadron collisions, at high transverse momenta. If, on the other hand, the prediction of the compact tetraquark Hamiltonian fails by few MeVs, as it could be the case considering the approximations involved, then a new $h$-tetraquark state will be observed, closely above $BK$. 

\vspace{.5cm}
{\emph{Note added}} --- Soon after the appearance of our preprint, we received a comment to our paper~\cite{wang,private}, arguing that the square root behavior shown in Fig.~\ref{fit_fesh} is a simple phase space factor unrelated to the underlying dynamics.  As already commented in the text, since the total width of the state is the sum of the partial ones, the application of phase space arguments would lead to conclude that the largest contribution comes from the threshold with the largest detuning -- the opposite of the argument outlined here in Section~\ref{sec:hybrid}.

\section*{Acknowledgments}
We wish to thank L.~Maiani, V.~Riquer, A.~Ali, F.~Piccinini  and A.L.~Guerrieri for useful and fruitful discussions.  We also thank Z.~G.~Wang for the comments on the manuscript. This material is based upon work supported in part by the U.S. Department of Energy, Office of Science, Office of Nuclear Physics under contract DE-AC05-06OR23177.

\end{document}